\documentclass[12pt]{article}
\usepackage{amsmath}
\usepackage{amssymb}
\usepackage{amstext}
\usepackage{amscd}
\usepackage{amsfonts}
\newcommand{\be}{\begin{equation}}
\newcommand{\nd}{\noindent}
\newcommand{\ee}{\end{equation}}
\newcommand{\ben}{\begin{eqnarray}}
\newcommand{\een}{\end{eqnarray}}

\title{{\bf q-Fourier Transform and its inversion-problem}}

\author{M. C. Rocca$^{1,\,2}$ and A. Plastino$^1$ \\$^1$
Instituto de F\'{\i}sica
(IFLP-CCT-Conicet)\\Universidad Nacional de La Plata (UNLP)\\
C.C. 67 (1900) La Plata, Argentina\\$^2$ Departamento de F\'{\i}sica\\
Fac. de Ciencias Exactas, UNLP}

\date{December 1, 2011}

\begin{document}

\maketitle

\begin{abstract}

Tsallis' q-Fourier transform is not generally one-to-one. It is
shown here that, if we eliminate the requirement that $q$ be
fixed, and let it instead ``float", a simple extension of the
$F_q-$definition, this procedure restores the one-to-one character.


\end{abstract}

\newpage

\renewcommand{\theequation}{\arabic{section}.\arabic{equation}}

\section{Introduction}

Nonextensive statistical mechanics (NEXT) \cite{[1],[2],AP}, a
current generalization of the Boltzmann�Gibbs (BG) one, is
actively studied in diverse areas of Science. NEXT is based on a
nonadditive (though extensive \cite{[3]}) entropic information
measure characterized by the real index q (with q = 1 recovering
the standard BG entropy). It has been applied to variegated
systems such as cold atoms in dissipative optical lattices
\cite{[4]}, dusty plasmas \cite{[5]}, trapped ions \cite{[6]},
spinglasses \cite{[7]}, turbulence in the heliosheath \cite{[8]},
self-organized criticality \cite{[9]}, high-energy experiments at
LHC/CMS/CERN \cite{[10]} and RHIC/PHENIX/Brookhaven \cite{[11]},
low-dimensional dissipative maps \cite{[12]}, finance \cite{[13]},
galaxies \cite{AP1}, Fokker-Planck equation's applications
\cite{AP2}, etc.

NEXT can be advantageously expressed via q-generalizations of
standard mathematical concepts (the logarithm and exponential
functions, addition and multiplication, Fourier transform (FT) and
the Central Limit Theorem (CLT) \cite{tq2,tq4,FFF}).
 The q-Fourier transform $F_q$ exhibits the nice property of
 transforming q-Gaussians into q-Gaussians \cite{tq2}.
Recently, plane waves, and the representation of the Dirac delta
into plane waves have been also generalized
\cite{[15],[16],tq1,tq4}.

A serious problem afflicts $F_q$. It is not generally one-to-one.
A detailed example is discussed below. In this work we show that
by recourse to a rather simple but efficient stratagem that
consists in \begin{itemize} \item eliminating the requirement that
$q$ be fixed and instead \item let it``float", \end{itemize} one
restores the one-to-one character.

\section{Generalizing the q-Fourier transform}

We define, following \cite{tq2}, a q-Fourier transform of $f(x)\in
L^1(\mathbb{R})$, $f(x)\geq 0$ as
\[F(k,q)=[H(q-1)-H(q-2)]\times \]
\begin{equation}
\label{ep1.1}
\int\limits_{-\infty}^{\infty} f(x) \{1+i(1-q)kx[f(x)]^{(q-1)}\}^{\frac {1} {1-q}}
\;dx
\end{equation}
where $H(x)$ is the Heaviside step function.

\nd {\sf The only difference between this definition and that
given in \cite{tq2} is that $q$ is not fixed and varies within the
interval $[1,2)$}. Herein lies the hard-core of our presentation.
This simple change of perspective  makes it is easy to find the
inversion-formula for (\ref{ep1.1}) by recourse to  the inverse
Fourier transform
\begin{equation}
\label{ep1.2} f(x)=\frac {1}
{2\pi}\int\limits_{-\infty}^{\infty}\left[\lim_{\epsilon\rightarrow
0^+} \int\limits_1^2 F(k,q)\delta (q-1-\epsilon)\;dq\right]
e^{-ikx}\;dk.
\end{equation}
As a consequence,  we see that this q-Fourier transform is
one-to-one, unlike what happens in \cite{tq3},\cite{tq5}.    In
the next section we give an illustrative example.

\setcounter{equation}{0}

\section{Example}

As an illustration we discuss  the example given by  Hilhorst in
Ref. (\cite{tq4}). Let $f(x)$ be
\begin{equation}
\label{ep4.1}
f(x)=
\begin{cases}
\left(\frac {\lambda} {x}\right)^{\beta}\;;\; x\in[a,b]\;;\; 0<a<b\;;\;\lambda>0 \\
0\;;\;x\; \rm{outside}\; [a,b].
\end{cases}
\end{equation}
The corresponding q-Fourier transform is
\begin{equation}
\label{ep4.2} F(k,q)={\lambda}^{\beta} \int\limits_a^b
x^{-\beta}\{1+i(1-q)k{\lambda}^{\beta(q-1)}
x^{1-\beta(q-1)}\}^{\frac {1} {1-q}} \;dx.
\end{equation}
 Effecting the change of variables
\[y=x^{1-\beta(q-1)},\]
we have for (\ref{ep4.2})
\[F(k,q)=[H(q-1)-H(q-2)]\times\]
\begin{equation}
\label{ep4.3} \frac {{\lambda}^{\beta}} {1-\beta(q-1)}
\int\limits_{a^{1-\beta(q-1)}}^{b^{1-\beta(q-1)}} y^{\frac
{\beta(q-2)} {1-\beta(q-1)}}\{1+i(1-q)k{\lambda}^{\beta(q-1)}
y\}^{\frac {1} {1-q}} \;dy.
\end{equation}
Now, (\ref{ep4.3}) can be rewritten in the useful form
\[F(k,q)=[H(q-1)-H(q-2)]\times\]
\[\left\{\left\{H(q-1)-H\left[q-\left(1+\frac {1} {\beta}\right)\right]\right\}\right.\times\]
\[\frac {{\lambda}^{\beta}} {1-\beta(q-1)}
\int\limits_{a^{1-\beta(q-1)}}^{b^{1-\beta(q-1)}}
y^{-\frac {\beta(2-q)} {1-\beta(q-1)}}\{1+i(1-q)k{\lambda}^{\beta(q-1)}
y\}^{\frac {1} {1-q}}
\;dy+\]
\[\left\{H\left[q-\left(1+\frac {1} {\beta}\right)\right]-H(q-2)\right\}\times\]
\begin{equation}
\label{ep4.4} \left.\frac {{\lambda}^{\beta}} {\beta(q-1)-1}
\int\limits_{b^{1-\beta(q-1)}}^{a^{1-\beta(q-1)}} y^{\frac
{\beta(q-2)} {1-\beta(q-1)}}\{1+i(1-q)k{\lambda}^{\beta(q-1)}
y\}^{\frac {1} {1-q}} \;dy\right\}.
\end{equation}
Taking into account that the involved integrals are defined in a
finite interval, we can cast (\ref{ep4.4}) as
\[F(k,q)=[H(q-1)-H(q-2)]\times\]
\[\left\{\left\{H(q-1)-H\left[q-\left(1+\frac {1} {\beta}\right)\right]\right\}\right.\times\]
\[\frac {{\lambda}^{\beta}} {1-\beta(q-1)}\lim_{\epsilon\rightarrow 0^+}
\int\limits_{a^{1-\beta(q-1)}}^{b^{1-\beta(q-1)}}
y^{-\frac {\beta(2-q)} {1-\beta(q-1)}}\{1+i(1-q)(k+i\epsilon){\lambda}^{\beta(q-1)}
y\}^{\frac {1} {1-q}}
\;dy+\]
\[\left\{H\left[q-\left(1+\frac {1} {\beta}\right)\right]-H(q-2)\right\}\times\]
\begin{equation}
\label{ep4.5} \left.\frac {{\lambda}^{\beta}}
{\beta(q-1)-1}\lim_{\epsilon\rightarrow 0^+}
\int\limits_{b^{1-\beta(q-1)}}^{a^{1-\beta(q-1)}} y^{\frac
{\beta(q-2)}
{1-\beta(q-1)}}\{1+i(1-q)(k+i\epsilon){\lambda}^{\beta(q-1)}
y\}^{\frac {1} {1-q}} \;dy\right\}.
\end{equation}
We now use  results of the Integral's table  \cite{tt3} to
evaluate (\ref{ep4.5}) and get
\[\lim_{\epsilon\rightarrow 0^+}\int\limits_{a^{1-\beta(q-1)}}^{\infty}
y^{-\frac {\beta(2-q)} {1-\beta(q-1)}}\{1+i(1-q)(k+i\epsilon){\lambda}^{\beta(q-1)}
y\}^{\frac {1} {1-q}}
\;dy=\]
\[\frac {(q-1)[1-\beta(q-1)]a^{\frac {q-2} {q-1}}} {(2-q)
[(1-q)i(k+i0){\lambda}^{\beta}]^{\frac {1} {q-1}}}\times \]
\[F\left(\frac {1} {q-1},\frac {2-q} {(q-1)[1-\beta(q-1)]},
\frac {1} {q-1} + \frac {\beta(2-q)} {1-\beta(q-1)};\right.\]
\begin{equation}
\label{ep4.6} \left.-\frac {1}
{(1-q)i(k+i0){\lambda}^{\beta(q-1)}a^{1-\beta(q-1)}}\right),
\end{equation}
and
\[\lim_{\epsilon\rightarrow 0^+}\int\limits_0^{a^{1-\beta(q-1)}}
y^{\frac {\beta(2-q)} {\beta(q-1)-1}}\{1+i(1-q)(k+i\epsilon){\lambda}^{\beta(q-1)}
y\}^{\frac {1} {1-q}}
\;dy=\]
\[\frac {[\beta(q-1)-1]a^{1-\beta}} {\beta-1}\times \]
\[F\left(\frac {1} {q-1},\frac {\beta-1} {\beta(q-1)-1},
\frac {\beta q-2} {\beta(q-1)-1};\right.\]
\begin{equation}
\label{ep4.7}
\left.(q-1)i(k+i0){\lambda}^{\beta(q-1)}a^{1-\beta(q-1)}\right),
\end{equation}
where $F(a,b,c;z)$ is the hypergeometric function. Thus we obtain
for $F(k,q)$
\[F(k,q)=[H(q-1)-H(q-2)]\times\]
\[\left\{\left\{H(q-1)-H\left[q-\left(1+\frac {1} {\beta}\right)\right]\right\}\right.\times\]
\[\frac {(q-1){\lambda}^{\beta}} {(2-q)
[(1-q)i(k+i0){\lambda}^{\beta}]^{\frac {1} {q-1}}}\times \]
\[\left\{a^{\frac {q-2} {q-1}}F\left(\frac {1} {q-1},\frac {2-q} {(q-1)[1-\beta(q-1)]},
\frac {1} {q-1} + \frac {\beta(2-q)} {1-\beta(q-1)};\right.\right.\]
\[\left.\frac {1} {(q-1)i(k+i0){\lambda}^{\beta(q-1)}a^{1-\beta(q-1)}}\right)-\]
\[ b^{\frac {q-2} {q-1}}F\left(\frac {1} {q-1},\frac {2-q} {(q-1)[1-\beta(q-1)]},
\frac {1} {q-1} + \frac {\beta(2-q)} {1-\beta(q-1)};\right.\]
\[\left.\left.\frac {1} {(q-1)i(k+i0){\lambda}^{\beta(q-1)}b^{1-\beta(q-1)}}\right)\right\}+\]
\[\left\{H\left[q-\left(1+\frac {1} {\beta}\right)\right]-H(q-2)\right\}
\frac {{\lambda}^{\beta}} {\beta-1}\times\]
\[\left\{a^{1-\beta}F\left(\frac {1} {q-1},\frac {\beta-1} {\beta(q-1)-1},
\frac {\beta q-2} {\beta(q-1)-1};\right.\right.\]
\[\left.(q-1)i(k+i0){\lambda}^{\beta(q-1)}a^{1-\beta(q-1)}\right)-\]
\[b^{1-\beta}F\left(\frac {1} {q-1},\frac {\beta-1} {\beta(q-1)-1},
\frac {\beta q-2} {\beta(q-1)-1};\right.\]
\begin{equation}
\label{ep4.8}
\left.\left.\left.(q-1)i(k+i0){\lambda}^{\beta(q-1)}b^{1-\beta(q-1)}\right)\right\}\right\}.
\end{equation}
As we can see from (\ref{ep4.8}), $F(k,q)$ is dependent of $a$ and
$b$, and, as consequence, one-to-one as has been shown in Section
2. \vskip 3mm

\nd However, and this is the crucial issue, if we {\bf fix} $q$
and select $\beta=1/(q-1)$ (\ref{ep4.8}) simplifies and adopts the
appearance
\[F(k,q)={\lambda}^{\frac {1} {q-1}}\frac {q-1} {2-q}
 \left[H(q-1)-H(q-2)\right]\times\]
\[\left[a^{\frac {q-2} {q-1}} F\left(\frac {1} {q-1},\frac {2-q} {q-1},
\frac {2-q} {q-1}; (q-1)i(k+i0){\lambda}\right)\right.- \]
\begin{equation}
\label{ep4.9} \left.b^{\frac {q-2} {q-1}} F\left(\frac {1}
{q-1},\frac {2-q} {q-1}, \frac {2-q}
{q-1};(q-1)i(k+i0){\lambda}\right)\right].
\end{equation}
With the help of the result given in \cite{tt4} for
\[F(-a,b,b,-z)=(1+z)^a,\]
we obtain for (\ref{ep4.9}):
\[F(k,q)={\lambda}^{\frac {1} {q-1}}\frac {q-1} {2-q}\left[H(q-1)-H(q-2)\right]\]
\begin{equation}
\label{ep4.10} \left(a^{\frac {q-2} {q-1}}-b^{\frac {q-2}
{q-1}}\right) \left[1+(1-q)ik\lambda\right]^{\frac {1} {1-q}}.
\end{equation}
Using now the expression for $\lambda$ of \cite{tq4}, i.e.,
\[\lambda=\left[\left(\frac {q-1} {2-q}\right)
\left(a^{\frac {q-2} {q-1}}-b^{\frac {q-2} {q-1}}
\right)\right]^{1-q},\] we have, finally,
\begin{equation}
\label{ep4.11} F(k,q)=\left[H(q-1)-H(q-2)\right]
\left[1+(1-q)ik\lambda\right]^{\frac {1} {1-q}},
\end{equation}
which is the result given by Hilthorst in  \cite{tq4},
demonstrating that $F(k,q)$ is not one-to-one. As a conclusion we
can say that for fixed $q$ the q-Fourier transform is NOT
one-to-one. On the contrary, as we have shown in section 2, when q
is NOT fixed, the q-Fourier transform is indeed one-to-one.

\section*{Conclusions}

In the present communication we have discussed the NOT one-to-one
nature of the q-Fourier transform $F_q$. We have shown that, if we
eliminate the requirement that $q$ be fixed and let it``float"
instead, such simple extension of the $F_q-$definition  restores
the desired one-to-one character.

\vskip 4mm

\nd {\bf Acknowledments} The authors thank Prof. C. Tsallis for
having called our attention to the present problem.

\end{document}